# An Investigation of Data Privacy and Utility Preservation using KNN Classification as a Gauge

Kato Mivule[1] and Claude Turner PhD[2]
[1]mivulek0220@students.bowiestate.edu, [2]cturner@bowiestate.edu
Computer Science Department, Bowie State University, Bowie, MD, USA

**Abstract** – It is obligatory that organizations by law safeguard the privacy of individuals when handling datasets containing personal identifiable information (PII). Nevertheless, during the process of data privatization, the utility or usefulness of the privatized data diminishes. Yet achieving the optimal balance between data privacy and utility needs has been documented as an NP-hard challenge. In this study, we investigate data privacy and utility preservation using KNN machine learning classification as a gauge.

**Keywords:** Data Privacy Preservation, Data Utility, Machine Learning, KNN Classification.

## I. INTRODUCTION

During the process of data privatization, the utility or usefulness of the privatized data diminishes. Yet achieving the optimal balance between data privacy and utility needs has been documented as an NP-hard challenge [1] [2]. In this study, we investigate data privacy and utility preservation using KNN machine learning classification as a gauge. As Cynthia Dwork succinctly and aptly stated [6]:

> "Perfect privacy can be achieved by publishing nothing at all, but this has no utility; perfect utility can be obtained by publishing the data exactly as received, but this offers no privacy".

In this study, we investigate data privacy and utility preservation using KNN machine learning classification as a gauge [4].

*Noise addition:* is a data privacy perturbative method that adds a random value, usually selected from a normal distribution with zero mean and a very small standard deviation, to sensitive numerical attribute values to ensure privacy [3] [8]. The general expression of noise addition as defined:

$$X + \varepsilon = Z \qquad (1)$$

Where $X$ is the original numerical dataset and $\varepsilon$ is the set of random values (noise) with a distribution $e \sim N(0, \sigma^2)$ that is added to $X$, and finally $Z$ is the privatized dataset.

This work was supported in part by the U.S. Department of Education HBGI Grant.
Claude Turner, PhD is an Associate Professor of Computer Science and Director for the Center for Cyber Security and Emerging Technologies at Bowie State University. (E-mail: cturner@bowiestate.edu).
Kato Mivule is a doctoral candidate, Computer Science Department, Bowie State University. (E-mail: mivulek0220@students.bowiestate.edu).

*K Nearest Neighbors (KNN):* is a classification method that matches items in the test data to those in the training data by measuring the distance between the two items. Any *k* items that are closer to each other are then placed in the same class. The Euclidean distance is the normally used distance measure for KNN expressed as follows [5]:

$$distance(x, y) = \sqrt{\sum_{i=1}^{n}(x_i - y_i)^2} \qquad (2)$$

## II. METHODOLOGY

In the first stage of our approach, we apply a data privacy procedure, in this case, noise addition, on the Iris dataset for privacy [7]. The privatized Iris dataset is then sent to the KNN machine learning classifier for training and testing using 10 fold cross validation; the classification error is quantified. If the classification error is lower or equal to a threshold, then better utility might be achieved, otherwise, we adjust the data privacy parameters and re-classify the results.

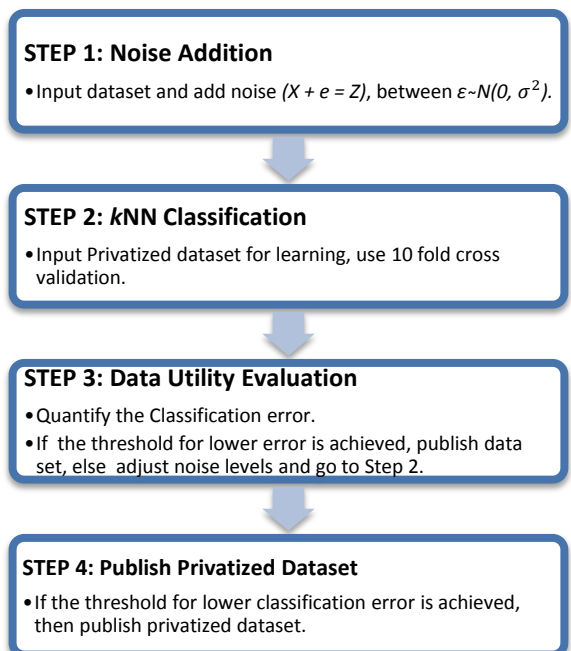

**STEP 1: Noise Addition**
- Input dataset and add noise *(X + e = Z)*, between $\varepsilon \sim N(0, \sigma^2)$.

**STEP 2: *k*NN Classification**
- Input Privatized dataset for learning, use 10 fold cross validation.

**STEP 3: Data Utility Evaluation**
- Quantify the Classification error.
- If the threshold for lower error is achieved, publish data set, else adjust noise levels and go to Step 2.

**STEP 4: Publish Privatized Dataset**
- If the threshold for lower classification error is achieved, then publish privatized dataset.

Fig.1: Data privacy and utility process using KNN classification as a gauge.

## III. EXPERIMENT

In our experiment, we used the Iris dataset from the UCI machine learning repository as our original dataset [9]. We then privatized the dataset by using the noise addition data privacy technique. We then used KNN classification and quantified the classification error. We adjusted the noise levels and run the privatized dataset through the KNN classifier after which we published the results. We used MATLAB for both noise addition and KNN classification.

## IV. RESULTS

As shown in our initial results, only 4 percent of records from the original Iris dataset were misclassified. When noise addition was chosen between the mean and standard deviation for the privatized dataset, 32 per cent of records got misclassified. However, when noise addition was reduced to mean = 0 and standard deviation = 0.1 for the privatized dataset, 26 percent of records got misclassified, a 6 point reduction in classification error.

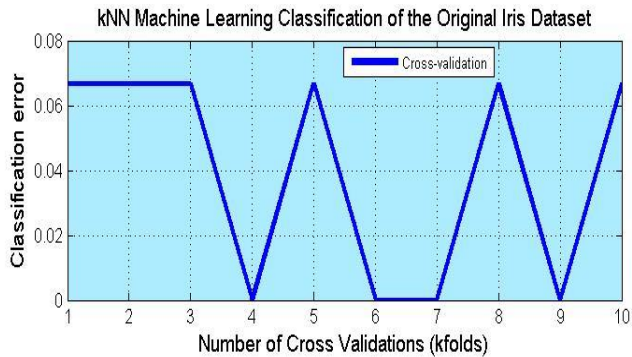

Fig 2: KNN classification of the original Iris dataset with classification error at 0.0400 (4 percent misclassified data)

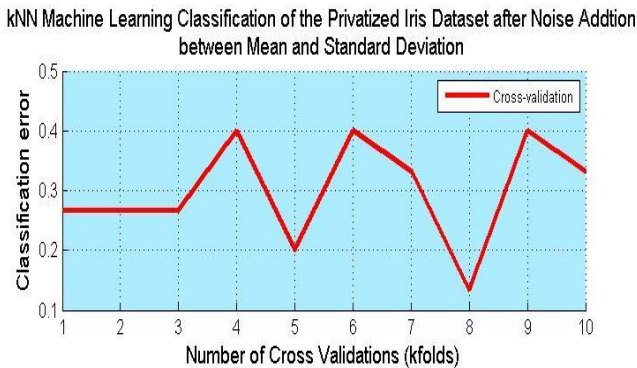

Fig 3: KNN classification of the privatized Iris dataset with noise addition between the mean and standard deviation.

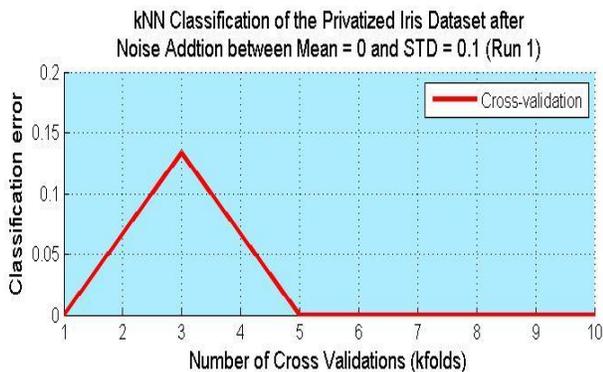

Fig 4: KNN classification of the privatized Iris dataset with reduced noise addtion between mean = 0 and standard deviation = 0.1

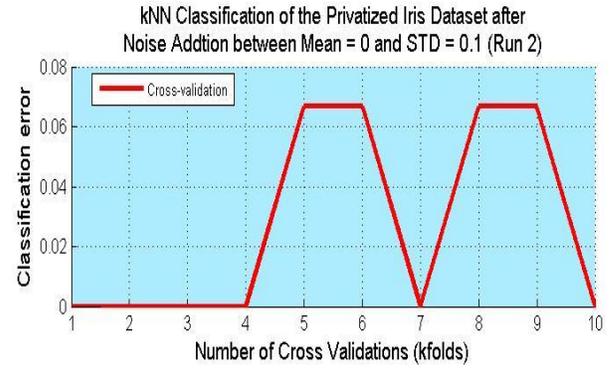

Fig 5: A second run of the *k*NN classification of the privatized Iris dataset with reduced noise addition between mean = 0 and standard deviation = 0.1.

## V. CONCLUSION AND DISCUSSION

The initial results from our investigation show that a reduction in noise levels does affect the classification error rate. However, this reduction in noise levels could lead to low risky privacy levels. Finding the optimal balance between data privacy and utility needs is still problematic.

### ACKNOWLEDGMENT

Special thanks to Dr. Claude Turner and the Computer Science Department at Bowie State University for making this work possible.